\documentclass[journal,twoside,web]{ieeecolor}
\usepackage{etoolbox}
\pdfoutput=1
\makeatletter
\@ifundefined{color@begingroup}%
{\let\color@begingroup\relax
	\let\color@endgroup\relax}{}%
\def\fix@ieeecolor@hbox#1{%
	\hbox{\color@begingroup#1\color@endgroup}}
\patchcmd\@makecaption{\hbox}{\fix@ieeecolor@hbox}{}{\FAILED}
\patchcmd\@makecaption{\hbox}{\fix@ieeecolor@hbox}{}{\FAILED}

\usepackage{generic}
\usepackage[justification=centering]{caption}
\usepackage{cite}
\usepackage{amssymb,amsfonts}
\usepackage{amsmath}
\usepackage{ntheorem}
\usepackage{algorithmic}
\usepackage{algorithm}
\usepackage{graphicx}
\usepackage{textcomp}
\usepackage{enumerate}
\usepackage[OT1]{fontenc}
\theorembodyfont{\texttt{}}
\theoremseparator{:}
\newtheorem{Definition}{Definition}
\newtheorem{Example}{Example}
\newtheorem{Remark}{Remark}
\newtheorem{Lemma}{Lemma}
\usepackage{hyperref}
\usepackage{color}
\hypersetup{
	colorlinks=true,
	linkcolor=blue,
	filecolor=blue,      
	urlcolor=blue,
	citecolor=cyan,
}
\hypersetup{
	colorlinks=true,
	linkcolor=blue,
	citecolor=blue,
}

\def\BibTeX{{\rm B\kern-.05em{\sc i\kern-.025em b}\kern-.08em
		T\kern-.1667em\lower.7ex\hbox{E}\kern-.125emX}}
\begin{document}
	\title{Enhanced \textsl{Q}-Learning Approach to Finite-Time Reachability with Maximum Probability for Probabilistic Boolean Control Networks}
	\author{Hongyue Fan, Jingjie Ni, Fangfei Li
	}
	\maketitle
	
	\begin{abstract}
		In this paper, we investigate the problem of controlling probabilistic Boolean control networks (PBCNs) to achieve reachability with maximum probability in the finite time horizon. We address three questions: 1) finding control policies that achieve reachability with maximum probability under fixed, and particularly, varied finite time horizon, 2) leveraging prior knowledge to solve question 1) with faster convergence speed in scenarios where time is a variable framework, and 3) proposing an enhanced \textsl{Q}-learning (\textsl{Q}L) method to efficiently address the aforementioned questions for large-scale PBCNs. For question 1), we demonstrate the applicability of \textsl{Q}L method on the finite-time reachability problem. For question 2), considering the possibility of varied time frames, we incorporate transfer learning (TL) technique to leverage prior knowledge and enhance convergence speed. For question 3), an enhanced model-free \textsl{Q}L approach that improves upon the traditional \textsl{Q}L algorithm by introducing memory-efficient modifications to address these issues in large-scale PBCNs effectively. Finally, we apply the proposed method to two examples: a small-scale PBCN and a large-scale PBCN, demonstrating the effectiveness of our approach.
	\end{abstract}
	
	
	\section{Introduction}
	\label{sec:introduction}\IEEEPARstart{K}{auffman} \cite{kauffman1969metabolic} first introduced Boolean networks (BNs) as a mathematical framework for modeling the dynamic behavior of genetic regulatory networks. In a BN, genes are modeled as discrete state variables, commonly described using Boolean values (0 or 1) to denote their active or inactive states. To enhance the practical applicability of BNs in scenarios that require desired system behavior, the development of BNs to Boolean control networks has been pursued by incorporating control inputs. As understanding deepens, researchers gradually realized the inherent limitations of BNs, such as their inability to effectively capture the intrinsic stochasticity that exists in biological, social, and engineering systems. Driven by the demand for more accurate and realistic modeling, Shmulevich et al. \cite{shmulevich2002probabilistic} proposed probabilistic Boolean networks, which expand the capabilities of BNs by introducing probabilistic elements, leading to a more faithful representation of system dynamics, such as genetic regulatory networks. Control inputs are commonly employed to manipulate the behavior of probabilistic Boolean networks, leading to the extensive investigation of probabilistic Boolean control networks (PBCNs). PBCNs find applications in various domains. Some of the application scenarios of PBCNs include: biological systems \cite{ma2008probabilistic}, manufacturing engineering systems \cite{rivera2019multiple}, credit default data \cite{6923626}, social networks \cite{li2014inferring}, etc. To date, numerous significant research topics related to probabilistic Boolean networks and PBCNs have been extensively discussed and investigated by scholars, including stability \cite{8688646}, reachability \cite{Zhao:2014aa}, optimal control \cite{9146736}, stabilization \cite{8700310}, and so on.
	
	The reachability remains a key research problem in the field of control with significant practical implications. For instance, when modeling the mechanism of drug action using PBCNs, the analysis of reachability in PBCNs can be utilized to investigate the drug's mode of action, assess its efficacy and evaluate potential adverse effects \cite{8585043}. H. Li et al. \cite{LI20122917} investigated the reachability and controllability of switched Boolean control networks using the semi-tensor product method. Y. Liu et al. \cite{LIU2015340} constructed the controllability matrix based on a new operator to investigate the controllability and reachability of PBCNs with forbidden states. Furthermore, an algorithm has been established to identify an optimal control policy that achieves reachability in the shortest possible time \cite{Zhao:2014aa}. However, most existing research on the reachability problem in PBCNs has relied on model-based approaches such as the semi-tensor product method. The semi-tensor product method is not applicable in model-free scenarios and the matrix calculations involved in these approaches can be highly complex when dealing the reachability for PBSNs. Therefore, a reinforcement learning-based approach is employed for the control problems for Boolean control networks (or PBCNs). Reinforcement learning offers a model-free framework, such as $Q$-learning ($Q$L) \cite{watkins1992q}, to solve some control problems modeled as Markov decision process \cite{puterman2014markov}. It is worth noting that researchers have successfully explored the use of $Q$L to address relevant control problems in PBCNs \cite{acernese2020reinforcement, zhou2022cluster, li2023edge}.
	
	Despite extensive research on the reachability of PBCNs, there are still unresolved issues in this field, primarily manifested in the following three aspects. Firstly, the existing studies on finite-time reachability of PBCNs, have focused on the case where the finite time $T$ is a fixed constant, without considering the variability of $T$. Due to the significant cost associated with long-term treatment, control design over an infinite horizon is often impractical \cite{ching2007optimal}. Therefore, investigating the finite-time reachability problem in PBCNs is of paramount importance. In \cite{zhou2019set}, Zhou et al. designed a random logic dynamical system and defined reachability for probabilistic Boolean networks within a finite time. Authors in \cite{pan2023finite} investigated the finite-time set reachability of PBCNs based on set reachability and parallel extension in the algebraic state space representation. However, their research on finite-time analysis are limited to the case where the finite time is a constant. In reality, the time requirements in the problem of finite-time probability reachability are often not precise time points but rather given time intervals. This is because time fluctuates due to environmental changes or variations in task objectives, and the ability to respond to changing conditions in real-time is crucial in practical environments. Therefore, considering more diverse finite-time scenarios, such as when the finite time is variable, is of significant importance. 
	
	Secondly, the existing research on reachability in PBCNs does not take into account the transferability of the model in control strategies. In fact, transferability has significant implications in practical applications. It can improve system efficiency and performance, reduce data requirements and costs, and enhance model generalization and adaptability. Inspired by \cite{CHEN2020119866}, we integrate transfer learning (TL) to solve the reachability with maximum probability for PBCNs in different time frames. TL involves utilizing knowledge from one or more related tasks as an additional source of information, beyond standard training data \cite{torrey2010transfer}. Incorporating TL allows PBCNs to achieve reachability within a finite time frame with faster convergence rates, which has significant practical implications.
	
	Thirdly, the exponential increase in states leads to higher complexity for large-scale PBCNs. As a result, existing research primarily focuses on small-scale PBCNs, with limited methods proposed for large-scale PBCNs. However, in practical applications such as genetic regulatory networks, manufacturing systems, intelligent transportation systems, and financial markets, the prevalence of large-scale and complex PBCNs are even more prominent. Scholars have proposed using reinforcement learning methods to study large-scale PBCNs, with commonly used approaches including the Deep $Q$-Network algorithm \cite{YERUDKAR2023374}. However, the algorithm does not guarantee convergence, while the $Q$L algorithm does. Additionally, the training and tuning of the Deep $Q$-Network are relatively complex. However, applying the $Q$L method to large-scale PBCNs may encounter issues such as excessive memory usage. Although \cite{peng2023qlearning} proposes an improved $Q$L algorithm to address the optimal false data injection attack problem in large-scale PBCNs, we note that \cite{peng2023qlearning} differs significantly from our problem. \cite{peng2023qlearning} investigates the maximum disruption of system performance by selecting the attack time points against large-scale PBCNs. Therefore, the reachability issue of large-scale PBCNs remains a challenging problem that requires further investigation. Furthermore,  the previously mentioned finite time $T$ being a variable, and the transferability of reachability results pose additional challenges for the reachability of large-scale PBCNs.
	
	Taking into account the aforementioned issues, our objective is to design a model-free approach to achieve reachability in finite time for PBCNs with maximum probability, including large-scale ones, within a framework that allows for time variability. Specifically, we aim to address three main challenges, mentioned in the abstract. The main contributions of this paper are as follows: 
	
	\begin{enumerate}
		\item Regarding question 1), we employ Algorithm 1 to investigate the problem of reachability with maximum probability for PBCNs, under a finite
		time horizon. In comparison to \cite{yang2022reachability}, our research considers a more diverse range of finite-time scenarios. Specifically, we not only consider the case when the finite time $T$ is a constant but also place particular emphasis on situations where $T$ is subject to a dynamic framework, such as following a certain distribution.
		\item In question 2), we consider the transferability of the model and incorporate TL into our Algorithm 2, leveraging existing knowledge from related tasks to improve convergence speed in the presence of varied time.
		\item  Compared to traditional $Q$L methods in \cite{li2023edge}, to address question 1) and 2)
		in large-scale PBCNs effectively, we propose an enhanced
		model-free $Q$L approach that improves upon
		the traditional $Q$L algorithm by introducing memory-efficient modifications. Therefore, Algorithm 3 and Algorithm 4 are proposed. This approach effectively reduces memory requirements.
	\end{enumerate}
	
	The paper is organized as follows. In Section II, we present the notations used in this paper and introduce the system model of PBCNs. Then, we further outline the problems addressed in our study. In Section III, the concept of reinforcement learning is introduced. In Section IV, we propose the method of enhanced $Q$L and investigate the problem of reachability with maximum probability for PBCNs within a finite time using the enhanced $Q$L approach. We address two scenarios: when time is constant and when time follows a distribution. Moreover, TL is incorporated into our methodology. This approach can be more effectively utilized in large-scale PBCNs. In Section V, the proposed method is validated using both small-scale and large-scale systems to assess its effectiveness. Section VI  is a brief conclusion.
	
	\section{ Problem Formulation}
	In this section, we first provide the definitions of the notations used throughout the subsequent texts. Then, we introduce our system model named PBCNs. Finally, we present the problem under investigation in this paper.
	\subsection{Notations}
	\begin{enumerate}
		\item $\mathbb{R}$, $\mathbb{Z}_{+}$, \text{and} $\mathbb{Z}_{+0}$ denote the sets of real numbers, positive integers, and nonnegative integers, respectively.
		\item $\mathcal{B}:=\{0,1\}\text{, and }\mathcal{B}^n:=\underbrace{\mathcal{B}\times\ldots\times\mathcal{B}}_n.$
		\item The basic logical operators, namely Negation,
		And, Or, are denoted by $\lnot,\land,\lor,$ respectively.
		\item $\mathbb{E}[\cdot]$ is the expected value operator.
		\item $\operatorname{var}[\cdot]$ represents the variance.
		\item $\mathcal{N}(\mu,\sigma^{2})$ means a normal distribution with mean $\mu$ and variance $\sigma^{2}$.
		\item $\mathbf{Pr}\{A\:|B\}$ represents the probability of event $A$ occurring given that event $B$ has occurred.
		\item  $||\cdot||$ refers to the $\mathcal{L}_2$-norm.
	\end{enumerate}
	
	\subsection{Probabilistic Boolean Control Networks}
	A PBCN with $n$ nodes and $m$ control inputs is defined as
	\begin{equation}
		\mathcal{X}^i(t+1)=f_i(\mathcal{U}(t),\mathcal{X}(t)),i=1,\ldots,n,\label{456}
	\end{equation}
	where $\mathcal{X}(t)=(\mathcal{X}^1(t),\ldots,\mathcal{X}^n(t))$ and $\mathcal{U}(t)=(\mathcal{U}^1(t),\ldots,\mathcal{U}^m(t))$ denote the $n$-dimensional state variable and $m$-dimensional control input variable at time $t\in\mathbb{Z}_{+0}$, in $\mathcal{B}^{n}$ and $\mathcal{B}^{m}$; $f_{i}:\mathcal{B}^{n+m}\rightarrow\mathcal{B},i=1,\ldots,n$ are logical functions.
	$f_{i}\in\mathcal{F}_{i}=\{f_{i}^{1},f_{i}^{2},\ldots,f_{i}^{l_{i}}\}$ is chosen randomly with probability $\{\mathsf{P}_{i}^{1},\mathsf{P}_{i}^{2},\ldots,\mathsf{P}_{i}^{l_i}\}$, where $\sum_{j=1}^{l_i}\mathsf{P}_i^j=1$ and $\mathsf{P}_{i}^{j}\geq0$.
	\begin{Definition}[\cite{7287128}]
		PBCNs is called reachable from $\mathcal{X}_0$ to $\mathcal{X}_d$ with a probability of $\omega$ within time $T$, only if there exist a sequence of controls $\{\mathcal{U}(0),\mathcal{U}(1),\cdots,\mathcal{U}(t-1)\},$ such that
		\begin{equation}
			\begin{aligned}
				\sum_{t=0}^{T}\mathbf{Pr}\{\mathcal{X}(t)=\mathcal{X}_d|\mathcal{X}(0)=\mathcal{X}_0,\mathcal{X}(t')\neq\mathcal{X}_d,t'=1, ...,\\  t-1, \mathcal{U}(0),\cdots,\mathcal{U}(t'-1)\}\geq\omega,
			\end{aligned}
		\end{equation}
		where $\omega$ refers to the probability of reaching the desired state of PBCNs within a finite time $T$.
	\end{Definition}
	\subsection{Problem Statements}
	In this subsection, we provide a concise summary of the problems addressed in this paper and highlight the challenges associated with studying these issues.
	\subsubsection{Problem 1}
	\label{P1}
	How to maximize the probability $\omega$ of reaching the target state in PBCNs when $T$ is a constant or a distribution?
	
	Addressing this problem poses a challenge due to the uncertainty introduced by the varied time frame. It requires the development of control strategies that optimize the system's behavior, aiming to achieve the highest probability of reaching the desired state within the specified time frame.
	\subsubsection{Problem 2}
	Considering the potential variation in time $T$ based on Problem 1, how can we utilize the training results of $T$ to accelerate the convergence of the results of $T+a$, where $a\in\mathbb{Z}_{+}$ and $a<<T$?\label{P2}
	\begin{Remark}
		$a<<T$ refers to the fact that the changes in time are minor.
	\end{Remark}
	
	Solving this problem presents challenges as it involves utilizing the knowledge gained from previous learning experiences to accelerate the convergence of the algorithm towards the results of a slightly extended time frame. This requires developing innovative methods that leverage TL and exploit the similarities between different time frames.
	\subsubsection{Problem 3}
	How to address the aforementioned problems for large-scale PBCNs?\label{P3}
	
	Large-scale PBCNs involve increased complexity and computational burden, requiring the development of efficient algorithms to overcome memory limitations, maintain convergence, and address the aforementioned problems effectively.
	\section{Preliminaries}
	In this section, we discuss the utilization of Markov decision processs and $Q$L in the field of reinforcement learning.
	\subsection{Reinforcement Learning: Markov Decision Process}
	Markov decision process is a theoretical framework in reinforcement learning that facilitates the achievement of objectives through interactive learning. We consider an Markov decision process which is defined by a tuple $(\mathbf{X},\mathbf{U},\mathbf{P},\mathbf{R},\gamma)$, where $\mathbf{X}$ is the set of states, $\mathbf{U}$ is the set of actions, $\mathbf{P}:\mathbf{X}\times\mathbf{U}\times\mathbf{X}\to[\mathbf{0},\mathbf{1}]$ is
	the function of state-transition probabilities that describes, for each state $\mathcal{X}_{t}\in\mathbf{X}$ and action $\mathcal{U}_{t}\in\mathbf{U}$, the conditional probability $\mathbf{P}_{\mathcal{X}_{t},\mathcal{X}_{t+1}}^{\mathcal{U}_{t}}=\mathbf{Pr}\{\mathcal{X}_{t+1}|\mathcal{X}_{t},\mathcal{U}_{t}\}$ of transitioning from $\mathcal{X}_{t}$ to $\mathcal{X}_{t+1}$, when $\mathcal{U}_t$ is taken. $\mathbf{R}:\mathbf{X}\times\mathbf{U}\times\mathbf{X}\to\mathbb{R}$ represents a reward function. The expected reward obtained after transitioning to state $\mathcal{X}_{t+1}$ is denoted as $\mathbf{R}_{\mathcal{X}_{t},\mathcal{X}_{t+1}}^{\mathcal{U}_{t}}=\mathbb{E}[r_{t+1}|\mathcal{X}_{t},\mathcal{U}_{t}]$, where $r_{t+1}=r_{t+1}(\mathcal{X}_{t},\mathcal{U}_{t},\mathcal{X}_{t+1})$. Here, $\begin{aligned}\gamma\in(0,1]\end{aligned}$ is a discount factor.
	
	We utilize the following formula to represent the return. The objective of the agent is to maximize the cumulative reward. 
	\begin{equation}
		G_t\doteq\sum_{k=t+1}^T\gamma^{k-t-1}r_k.
	\end{equation}
	
	We also define the value of taking action $\mathcal{U}_t$ in state $\mathcal{X}_t$ under a policy $\pi$, denoted $q_{\pi}(\mathcal{X}_{t},\mathcal{U}_{t})$, as the expected return starting from $\mathcal{X}_t$, taking the action $\mathcal{U}_t$, and thereafter following policy $\pi$:
	\begin{equation}
		q_{\pi}(\mathcal{X}_{t},\mathcal{U}_{t})\doteq\mathbb{E}_{\pi}[G_{t}|\mathcal{X}_{t},\mathcal{U}_{t}]=\mathbb{E}_{\pi}\bigg[\sum_{k=0}^{T}\gamma^{k}r_{t+k+1}\bigg|\mathcal{X}_{t},\mathcal{U}_{t}\bigg].
	\end{equation}
	
	The Bellman optimality equation provides a recursive definition of the optimal value function. By solving the Bellman optimality equation, we can determine the optimal value function.
	
	\begin{equation}
		q_{*}(\mathcal{X}_{t},\mathcal{U}_{t}) =\sum_{\mathcal{X}\in\mathbf{X},r}p(\mathcal{X},r|\mathcal{X}_{t},\mathcal{U}_{t})\Big[\mathbf{R}_{\mathcal{X}_{t},\mathcal{X}}^{\mathcal{U}_{t}}+\gamma\operatorname*{max}_{\mathcal{U}\in\mathbf{U}}q_{*}(\mathcal{X},\mathcal{U})\Big].
	\end{equation}
	
	Under the optimal policy $\pi^{*}$, the value function of each state reaches its maximum value, and the agent selects the corresponding optimal actions to achieve the goal of maximizing the cumulative reward. Markov decision process provides a framework for describing sequential decision problems and reinforcement learning is one of the approaches to solving Markov decision process problems, where $Q$L is a common algorithm in reinforcement learning. Next, we introduce $Q$L.
	\subsection{Reinforcement Learning: Q-learning}
	$Q$L is an off-policy temporal-difference (TD) control
	algorithm used to learn optimal decision-making in unknown environments, defined by
	\begin{equation}
		Q(\mathcal{X}_{t},\mathcal{U}_{t})\leftarrow Q(\mathcal{X}_{t},\mathcal{U}_{t})+\alpha_{t}TDE_{t+1},
	\end{equation}
	\begin{equation}
		TDE_{t+1} = r_{t+1}+\gamma\max_{\mathcal{U}\in\mathbf{U}}Q(\mathcal{X}_{t+1},\mathcal{U})-Q(\mathcal{X}_{t},\mathcal{U}_{t}),
	\end{equation}
	where $TDE_{t+1}$ is known as temporal-difference error, $\alpha_{t}$ is the learning rate and $0 < \alpha_{t} \leq 1$.
	
	In this case, the learned action-value function $Q$ directly approximates the optimal action-value function $q_{*}$ \cite{sutton2018reinforcement}. The agent's behavior during training is determined by an $\epsilon$-greedy policy given as follows:
	\begin{equation}
		\mathcal{U}_t=\begin{cases}\arg\max\limits_{\mathcal{U}\in\mathbf{U}}Q_t\left(\mathcal{X}_t,\mathcal{U}\right),&\text{if}\text{		} \text{rand}>\epsilon;\\\operatorname{rand}(\mathbf{U}),&\text{otherwise,}\end{cases}\label{0905}
	\end{equation}
	where \text{rand} is a random number generated from the interval  [0, 1], $\operatorname{rand}(\mathbf{U})$ is  an action randomly selected
	from $\mathbf{U}$ and $0 < \epsilon \leq 1$.
	
	\section{Finite Time Reachability with Maximum Probability for PBCNs Using Enhanced QL}
	In this section, we will first introduce the Markov decision process framework and propose the so called enhanced $Q$L approach.  Next, we develop four algorithms to investigate the problem of achieving the maximum probability of reaching the desired state in finite time for PBCNs. These algorithms include traditional $Q$L, enhanced $Q$L, traditional $Q$L with TL and enhanced $Q$L with TL. 
	\subsection{Markov Decision Process Framework}
	In our Markov decision process framework, the states $\mathcal{X}_t=\mathcal{X}(t):=(\mathcal{X}_t,t)$ and the actions $\mathcal{U}_t=\mathcal{U}(t)$ are selected using an $\epsilon$-greedy strategy as shown in Equation \eqref{0905}. Furthermore, we define the following reward signals:
	\begin{equation}
		r_{t}(\mathcal{X}_{t},\mathcal{U}_{t})=\begin{cases}1,&\mathcal{X}_{t}=(\mathcal{X}_{d},t),\:\forall t;\\\operatorname-1,&\mathcal{X}_{T} \neq (\mathcal{X}_{d},T);\\\operatorname0,&\text{else.} \end{cases}
	\end{equation}
	
	We define the goal-directed reward signal as mentioned before. A positive reward $1$ is given when the desired state is successfully reached for $t\leq T$, a negative reward $-1$ is given if the goal state is not reached at $t=T$, and a reward of $0$ is given otherwise. This reward signal design aims to guide the agent in learning adaptive behaviors. By appropriately setting the reward signals, we can incentivize the agent to learn and optimize its behavior towards the desired state.
	In the subsequent algorithm, we also set $\gamma$ = 1 for all iterations. This is because, in reinforcement learning, $\gamma$ determines the discounting factor for future rewards. When $\gamma$ is set to 1, it means that future rewards are fully preserved without any discounting. This setting ensures that all future rewards are thoroughly considered during the learning process, thereby maximizing the probability of reaching the desired system state. Under this specific definition of $\gamma$, $Q$L is also feasible, as stated in the following lemma.
	\begin{Lemma}[\cite{jaakkola1993convergence}]
		$Q_{t}(\mathcal{X}_{t},\mathcal{U}_{t})$ converges to the optimal values $Q^{*}(\mathcal{X}_{t},\mathcal{U}_{t})$ with probability one (w.p.1) under the following conditions:
		\begin{enumerate}
			\item $|\mathbf{X}|$ and $|\mathbf{A}|$ are finite.
			\item $\sum_{t=0}^\infty\alpha_t=\infty\text{ and }\sum_{t=0}^\infty\alpha_t^2<\infty$ uniformly w.p.1.
			\item $\operatorname{var}[r_t]$ is finite.
			\item If $\gamma=1$, all policies lead to a cost-free terminal state w.p.1.
		\end{enumerate}
	\end{Lemma}
	
	Our Markov decision process framework satisfies all the conditions stated in the above lemma.
	\subsection{Using Traditional QL to Solve \nameref{P1}}
	In this subsection, we employ the traditional $Q$L approach to address the problem of maximizing the probability of reaching a desired state in finite time for PBCNs, which leads to the introduction of Algorithm 1. We consider two scenarios:
	\begin{itemize}
		\item Case 1: When $T$ is a constant.
		\item Case 2: When $T$ follows a  specific distribution. 
	\end{itemize}
	
	For case 2, we consider a distribution with a cumulative distribution function $F(x)$ defined over the domain $x\in (a, b)$. We consider a partitioning of the domain into intervals $(a, n'), [n', n'+1), [n'+2, n'+3), ..., [n'+n'', b)$, where $n'$ and $n''$ are positive integers on the domain. By selecting several consecutive intervals from the above partition, we can divide the domain into several parts. Next, we establish the following rules:
	\begin{equation}
		T=\begin{cases}n^{\prime}-1,\mathsf{P}=\int_{a}^{n^{\prime}}F(x)dx;\\n^{\prime}+n,\mathsf{P}=\int_{n^{\prime}+n}^{n^{\prime}+n+1}F(x)dx,\forall n=0,1,\cdots,n^{\prime\prime}-1;\\n^{\prime}+n^{\prime\prime},\mathsf{P}=\int_{n^{\prime}+n^{\prime\prime}}^{b}F(x)dx.\end{cases}\label{1119}
	\end{equation}
	Therefore, for each distribution, we can discretize the continuous formulation using Equation \eqref{1119}, making the $Q$L algorithm still applicable.
	
	We transform the reachability problem of PBCNs into the $Q$L framework, where $Q$L continuously updates the $Q$-table to learn the optimal policy for different actions in different states, thus achieving the maximum probability of reaching the target. The exploration-exploitation strategy in the $Q$L ensures that the system explores unknown territories while exploiting known information during the learning process. This approach is a model-free reinforcement learning method. 
	
	\begin{algorithm}[!h]
		\caption{Time-Constrained Reachability with Maximum Probability for PBCNs Using Traditional $Q$L}
		\label{alg:AOS}
		\renewcommand{\algorithmicrequire}{\textbf{Input:}}
		\renewcommand{\algorithmicensure}{\textbf{Output:}}
		
		\begin{algorithmic}[1]
			\REQUIRE $\omega, \gamma, \beta, N, T, \mathcal{X}_{0}, \mathcal{X}_d$  
			\ENSURE $\ Q^*(\mathcal{X}_t,\mathcal{U}_t)$   
			
			\STATE  Initialize: $Q_{0}(\mathcal{X}_{t},\mathcal{U}_{t})\leftarrow0$, $\forall \mathcal{X}_{t}\in\mathcal{B}^n,\,\forall \mathcal{U}_{t}\in\mathcal{B}^m$
			
			\FOR{$ep$ = 1, 2, \ldots, $N$}
			\STATE $t\leftarrow0,\,\mathcal{X}_t\leftarrow\mathcal{X}_{0}$
			\WHILE{$(t \leq T)\wedge(\mathcal{X}_{t}\neq\mathcal{X}_{d})$}
			\STATE $\mathcal{U}_t\leftarrow\epsilon\text{-greedy}(\epsilon,\mathcal{X}_t)$
			\STATE $t+1,\mathcal{X}_{t+1},r_{t+1}\leftarrow\operatorname{apply}(\mathcal{U}_t)$
			\STATE $\begin{aligned}&Q(\mathcal{X}_{t},\mathcal{U}_{t})\leftarrow\alpha_{ep}(r_{t+1}+\gamma\max_{\mathcal{U}}Q(\mathcal{X}_{t+1},\mathcal{U}_{t+1}))+\\&(1-\alpha_{ep})Q(\mathcal{X}_t,\mathcal{U}_{t})\end{aligned}$
			\STATE $t\leftarrow t+1$
			\ENDWHILE
			\STATE $Q^{*}(\mathcal{X}_{t},\mathcal{U}_{t})\leftarrow Q_{t}(\mathcal{X}_{t},\mathcal{U}_{t})$
			\ENDFOR
			\RETURN $\ Q^*(\mathcal{X}_t,\mathcal{U}_t)$
		\end{algorithmic}
	\end{algorithm}
	In the Algorithm 1, $\epsilon\text{-greedy}(\epsilon,\mathcal{X}_t)$ denotes the utilization of Equation \eqref{0905} for updating $\mathcal{U}_{t}$. $\operatorname{Apply}(\mathcal{U}_t)$ represents the utilization of the obtained $\mathcal{U}_{t}$ to update $t$, $\mathcal{X}_t$, and $r_t$.
	\subsection{Using Traditional QL Combined with TL to Solve \nameref{P2}}
	In this subsection, we propose Algorithm 2 which builds upon the concept of TL in conjunction with Algorithm 1. Through the integration and comparison of two TL methods, we are able to improve the convergence rate.
	
	We will introduce the concept of TL to investigate the problem of maximizing the probability of reaching a desired state in finite time for PBCNs. This concept primarily manifests in our modification of the $Q$-table initialization. In particular, if we aim to study the aforementioned problem within a time frame of $T+a$, we can utilize the $Q$-table obtained from the convergence of problem 1. By utilizing the existing knowledge, we enhance the convergence speed of the algorithm 2. 
	\begin{Remark}
		Based on the defined reward signal, we can determine that the value range of $G_t$ is within $[-1, 1]$. In this case, we set the initial $Q$-value for traditional $Q$L to be the midpoint of the range, which is 0.
	\end{Remark}
	
\begin{figure}[!h]
		{\includegraphics[width=\columnwidth]{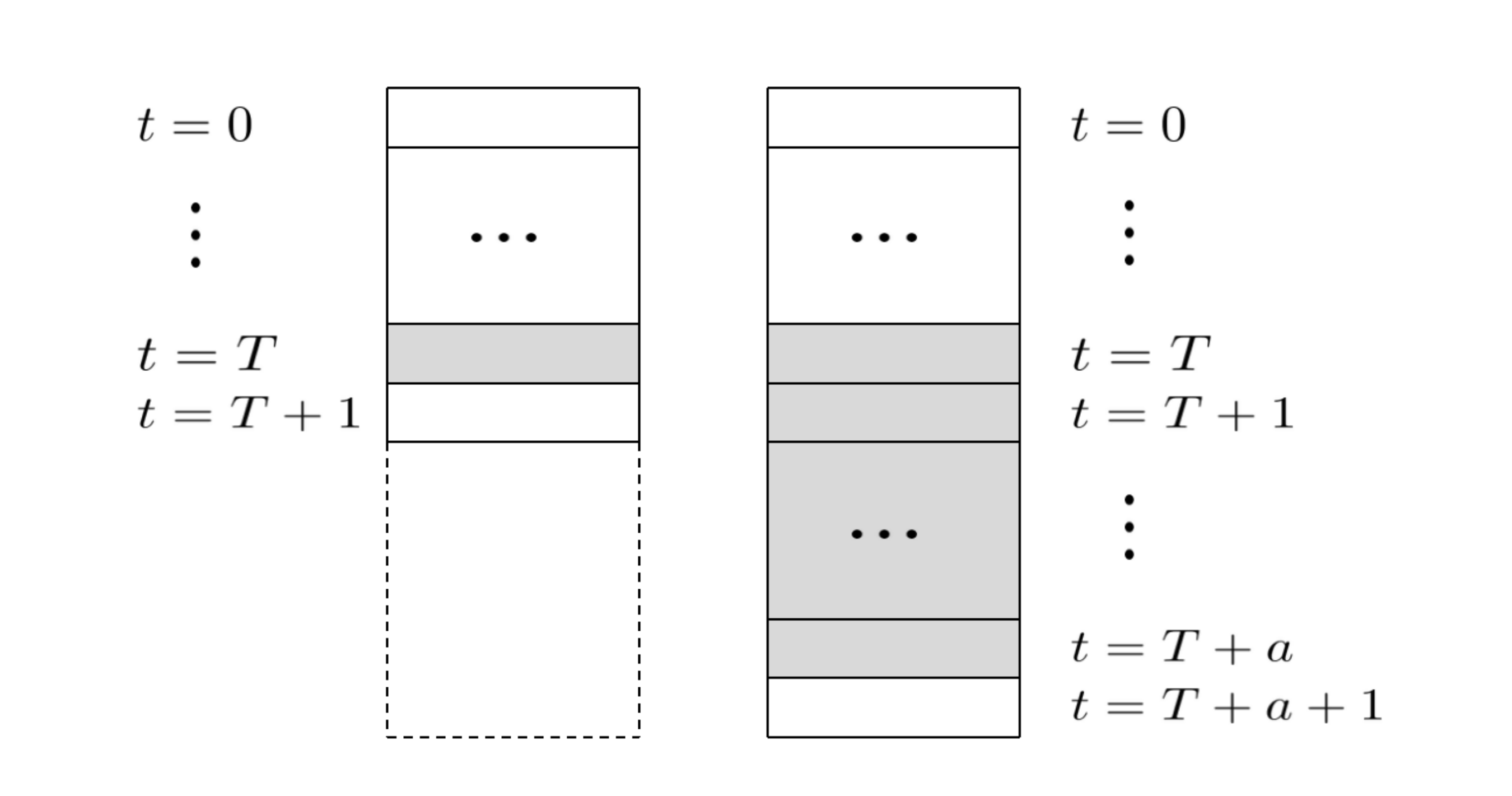}}
		\caption{Description of method 2 for TL.}
		\label{fig2}
	\end{figure}

	We propose the following two TL approaches. The settings of the two TL methods are as follows:
	\begin{itemize}
		\item Method 1: We pad the $Q$-table at time $T$ to ensure its dimensions are consistent with the $Q$-table at time $T+a$. Here, we use the initial value 0 of the training $Q$-table to fill it.
		\begin{equation}
			Q_{0}(\mathcal{X}_{t'},\mathcal{U}_{t'})\leftarrow
			0.
		\end{equation}
		This augmented $Q$-table serves as the initial $Q$-table for studying problem 2.
		\item Method 2: We duplicate the $Q^{*}(\mathcal{X}_{T},\mathcal{U}_{T})$ and assign them to $Q$-values at time $T+1, \cdots, T+a$.
		\begin{equation}
			Q_{0}(\mathcal{X}_{t'},\mathcal{U}_{t'})\leftarrow
			Q^{*}(\mathcal{X}_{T},\mathcal{U}_{T}), \forall t^{\prime}\in T+1, \cdots, T+a.
		\end{equation}
		
		This approach is taken to leverage more $Q$-values converged at time $T$ and utilize more prior knowledge. As shown in Figure \ref{fig2}, each $Q$-table at every time step consists of $2^n$ rows, with the gray region indicating the duplicated portion. Subsequently, this modified $Q$-table is utilized as the initial $Q$-table. 
	\end{itemize}
	
	Using the aforementioned approaches, we can obtain two types of initial $Q$-tables for training at time $T+a$. These initial Q-tables are as follows:
	\begin{equation}
		\begin{cases}
			Q_{0}(\mathcal{X}_{t},\mathcal{U}_{t})\leftarrow Q^{*}(\mathcal{X}_{t},\mathcal{U}_{t}),\\
			\mathcal{X}_{t}\:\text{represents the states encountered in\:} Q^{*}(\mathcal{X}_{t},\mathcal{U}_{t});\\
			Q_{0}(\mathcal{X}_{t'},\mathcal{U}_{t'})\leftarrow
			\text{by using Method 1 or 2,}\label{q}\\
			\mathcal{X}_{t'}\:\text{represents the states aren't encountered in\:} Q^{*}(\mathcal{X}_{t},\mathcal{U}_{t}),
		\end{cases}
	\end{equation}
	where $Q_{0}(\mathcal{X}_{t},\mathcal{U}_{t})$ represents the initial $Q$-table at time $T+a$. While $Q^{*}(\mathcal{X}_{t},\mathcal{U}_{t})$ represents the $Q$-table that has converged at time $T$ after augmenting its dimensions. Based on the aforementioned discussions, we propose Algorithm 2.
	\begin{algorithm}[!h]
		\caption{Time-Constrained Reachability with Maximum Probability for PBCNs Using Traditional $Q$L Combined with TL}
		\label{alg:AOS}
		\renewcommand{\algorithmicrequire}{\textbf{Input:}}
		\renewcommand{\algorithmicensure}{\textbf{Output:}}
		
		\begin{algorithmic}[1]
			\REQUIRE $\omega, \gamma, \beta, N, T, a, \mathcal{X}_{0}, \mathcal{X}_d,\ Q^{*}(\mathcal{X}_t,\mathcal{U}_t) $  
			\ENSURE $\ Q^{*'}(\mathcal{X}_t,\mathcal{U}_t)$   
			
			\STATE Applying Method 1 or Method 2 to process $\ Q^{*}(\mathcal{X}_t,\mathcal{U}_t)$.
			\STATE  Initialize: $Q_{0}(\mathcal{X}_{t},\mathcal{U}_{t})\leftarrow Q^{*}(\mathcal{X}_{t},\mathcal{U}_{t})$, $\forall \mathcal{X}_{t}\in\mathcal{B}^n,\,\forall \mathcal{U}_{t}\in\mathcal{B}^m$
			
			\FOR{$ep$ = 1, 2, \ldots, $N$}
			\STATE $t\leftarrow0,\,\mathcal{X}_t\leftarrow\mathcal{X}_{0}$
			\WHILE{$(t \leq T)\wedge(\mathcal{X}_{t}\neq\mathcal{X}_{d})$}
			\STATE $\mathcal{U}_t\leftarrow\epsilon\text{-greedy}(\epsilon,\mathcal{X}_t)$
			\STATE $t+1,\mathcal{X}_{t+1},r_{t+1}\leftarrow\operatorname{apply}(\mathcal{U}_t)$
			\STATE $\begin{aligned}&Q(\mathcal{X}_{t},\mathcal{U}_{t})\leftarrow\alpha_{ep}(r_{t+1}+\gamma\max_{\mathcal{U}}Q(\mathcal{X}_{t+1},\mathcal{U}_{t+1}))+\\&(1-\alpha_{ep})Q(\mathcal{X}_t,\mathcal{U}_{t})\end{aligned}$
			\STATE $t\leftarrow t+1$
			\ENDWHILE
			\STATE $Q^{*'}(\mathcal{X}_{t},\mathcal{U}_{t})\leftarrow Q_{t}(\mathcal{X}_{t},\mathcal{U}_{t})$
			\ENDFOR
			\RETURN $\ Q^{*'}(\mathcal{X}_t,\mathcal{U}_t)$
		\end{algorithmic}
	\end{algorithm}
	
	\subsection{Using Enhanced QL to Solve \nameref{P3}}
	\subsubsection{Enhanced Q-learning}
	\label{EQ}
	In our Markov decision process, a PBCN with $n$ nodes has $2^n \times t$ states, which grows exponentially with $n$. However, in reality, a significant portion of these states remains unvisited, and we don't need to study them, thus eliminating the need to allocate $Q$-table positions for them. Therefore, we propose an improvement to the traditional $Q$L algorithm to reduce memory requirements and significantly enhance the utilization of the $Q$-table.
	
	Inspired by \cite{peng2023qlearning}, we have modified the traditional $Q$L algorithm to enhance its efficiency. Instead of initializing the 
	$Q$-table with a fixed size, we incrementally increase its dimensions when encountering new state-time pairs. 
	
	This approach significantly reduces memory requirements and optimizes storage efficiency. In fact, in the problem of finite-time $T$ maximum probability reachability that we are investigating, PBCNs with $n$ nodes have $2^{n} \times (T+1)$ possible states. However, in practice, the number of visited states is significantly smaller than this total, as Equation \eqref{456} used to update $\mathcal{X}_t$ is discrete, leading to many states remaining untraversed. Therefore, traditional $Q$L suffers from excessive memory usage, while our approach reduces memory requirements by storing only the necessary state-time pairs, thus optimizing storage efficiency.
	
\begin{figure}[!h]
		{\includegraphics[width=\columnwidth]{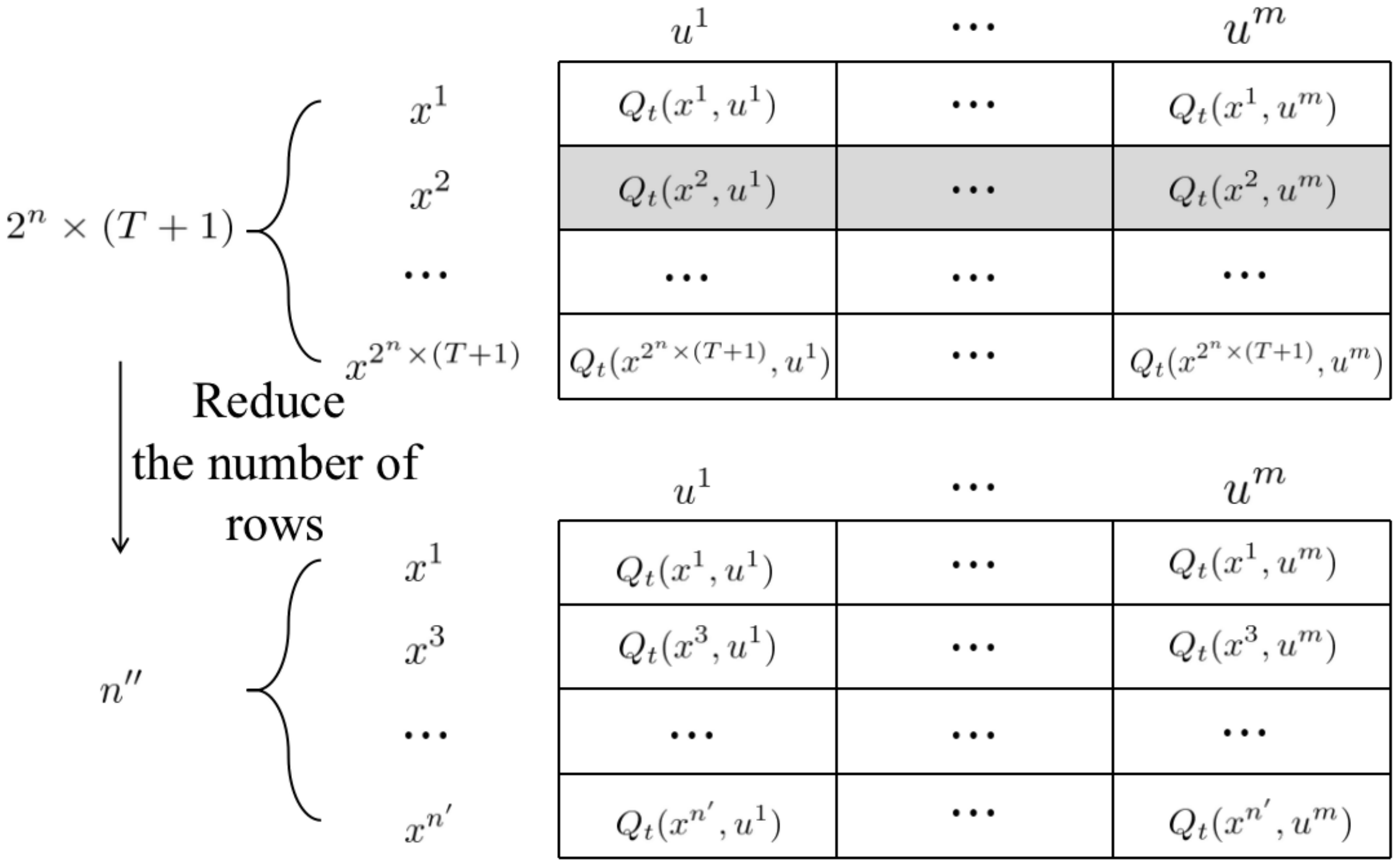}}
		\caption{Comparison between traditional and enhanced $Q$L in terms of $Q$-table handling.}
		\label{fig1}
	\end{figure}

	As shown in Figure \ref{fig1}, we present the $Q$-table of a PBCN with $n$ nodes and $m$ action choices under both traditional and enhanced $Q$L approaches in the problem of finite-time maximum probability reachability. In the Figure, $x^{n}$ represents the $n$-th state, and $u^{m}$ represents the $m$-th action. In traditional $Q$L, a fixed-size $Q$-table of dimensions $[2^n \times (T+1)]\times m$ is maintained. However, this includes states that have not been visited, as exemplified by state $x^2$ in Figure \ref{fig1}. The actual number of states encountered is only $n''$. Thus, the shaded area in Figure \ref{fig1} represents the ineffective memory utilization of the traditional $Q$-table.
	To address this inefficiency, we no longer initialize the $Q$-table with $2^n \times (T+1)$ rows and $m$ columns. Instead, we fix the number of columns in the $Q$-table and initialize it with $0$ rows. As new states (eg. $x^{n'}$) are encountered, we dynamically add rows to the $Q$-table. This significantly improves the utilization of the $Q$-table.
	\subsubsection{Using Enhanced QL to Solve Problem 3}
	In this subsection, we present Algorithm 3 as an improvement to Algorithm 1 and use it to  address problem 3. Considering the memory requirements of traditional $Q$L for large-scale PBCNs, we utilize the Algorithm 3 approach to tackle the problem, where $ \mathcal{X}_{rec}$ is a matrix used to store the visited states. Using $ \mathcal{X}_{rec}$, we are able to determine if new states have been generated. If the state has been generated, we can locate it in the $Q$-table using $ \mathcal{X}_{rec}$ and update its $Q$-table accordingly. If it has not been generated, we must expand the $Q$-table by adding additionally rows.
	\begin{Remark}
		Algorithm 3 is equally applicable to the two scenarios mentioned in Algorithm 1.
	\end{Remark}
	\begin{algorithm}[!h]
		\caption{Time-Constrained Reachability with Maximum Probability for PBCNs Using Enhanced $Q$L}
		\label{alg:AOS}
		\renewcommand{\algorithmicrequire}{\textbf{Input:}}
		\renewcommand{\algorithmicensure}{\textbf{Output:}}
		
		\begin{algorithmic}[1]
			\REQUIRE $\omega, \gamma, N, T, \mathcal{X}_{0}, \mathcal{X}_d$  
			\ENSURE $\ Q^*(\mathcal{X}_t,\mathcal{U}_t), \mathcal{X}_{rec}$   
			
			\STATE  Initialize: $Q(\mathcal{X}_{t},\mathcal{U}_{t})\leftarrow$ an empty table, $\mathcal{X}_{rec}\leftarrow$ an empty table, $\forall \mathcal{X}_{t}\in\mathcal{B}^n,\,\forall \mathcal{U}_{t}\in\mathcal{B}^m$
			
			\FOR{$ep$ = 0, 1, \ldots, $N-1$}
			\STATE $t\leftarrow0,\,\mathcal{X}_t\leftarrow\mathcal{X}_{0}$
			\STATE Initialize $\mathcal{X}_{t} = [\mathcal{X}_{0}, 0] $
			\IF {$\mathcal{X}_{t} \,\text{not in the}\, \mathcal{X}_{rec}$}
			\STATE Add $\mathcal{X}_{t}$ into $\mathcal{X}_{rec}$ as a new row
			\STATE Expand $Q$-table by adding one additional row
			\STATE $Q^{*}(\mathcal{X}_{t},\mathcal{U}_{t})\leftarrow Q(\mathcal{X}_{t},\mathcal{U}_{t})$
			\ELSE
			\STATE $Q(\mathcal{X}_{t},\mathcal{U}_{t})\leftarrow Q^{*}(\mathcal{X}_{t},\mathcal{U}_{t})$
			\ENDIF
			
			\WHILE{$(t \leq T)\wedge(\mathcal{X}_{t}\neq\mathcal{X}_{d})\lor(t == 0)$}
			\STATE $\mathcal{U}_t\leftarrow\epsilon\text{-greedy}(\epsilon,\mathcal{X}_t)$
			\STATE $t+1,\mathcal{X}_{t+1},r_{t+1}\leftarrow\operatorname{apply}(\mathcal{U}_t)$
			\STATE $\mathcal{X}_{t+1} = [\mathcal{X}_{t+1}, t+1]$
			
			\IF {$\mathcal{X}_{t+1}\,\text{not in the}\, \mathcal{X}_{rec}$}
			\STATE Add $\mathcal{X}_{t+1}$ into $\mathcal{X}_{rec}$ as a new row
			\STATE Expand $Q$-table by adding one additional row
			\STATE $Q^{*}(\mathcal{X}_{t+1},\mathcal{U}_{t+1})\leftarrow Q(\mathcal{X}_{t+1},\mathcal{U}_{t+1})$
			\ELSE
			\STATE $Q(\mathcal{X}_{t+1},\mathcal{U}_{t+1})\leftarrow Q^{*}(\mathcal{X}_{t+1},\mathcal{U}_{t+1})$
			\ENDIF
			
			\IF {$((t-1) \neq T)\lor(\mathcal{X}_{t} = \mathcal{X}_{d})$}
			\STATE $\begin{aligned}&Q(\mathcal{X}_{t},\mathcal{U}_{t})\leftarrow\alpha_{ep}(r_{t+1}+\gamma\max_{\mathcal{U}}Q(\mathcal{X}_{t+1},\mathcal{U}_{t+1}))+\\&(1-\alpha_{ep})Q(\mathcal{X}_t,\mathcal{U}_{t})\end{aligned}$
			\ELSE
			\STATE
			$\begin{aligned}&Q(\mathcal{X}_{t},\mathcal{U}_{t})\leftarrow\alpha_{ep}\times r_{t+1}+(1-\alpha_{ep})Q(\mathcal{X}_t,\mathcal{U}_{t})\end{aligned}$
			\ENDIF
			
			\ENDWHILE
			\STATE $Q^{*}(\mathcal{X}_{t},\mathcal{U}_{t})\leftarrow Q(\mathcal{X}_{t},\mathcal{U}_{t})$
			\ENDFOR
			\RETURN $\ Q^*(\mathcal{X}_t,\mathcal{U}_t), \mathcal{X}_{rec}$
		\end{algorithmic}
	\end{algorithm}
	
	\subsection{Using Enhanced QL Combined with TL to Solve \nameref{P3}}
	Due to a slight adjustment in the reachable time of the PBCN model, we trained the enhanced $Q$L algorithm on the pre-adjustment PBCN model and obtained $Q^{*}$. In this subsection, we propose Algorithm 4 by incorporating the idea of TL on top of Algorithm 3 to accelerate the convergence rate of the algorithm. In comparison to Algorithm 3, we make modifications in the initialization by utilizing Equation \eqref{q} and set $\mathcal{X}_{rec}^{'}$ as the initial value of $\mathcal{X}_{rec}$.
	\begin{algorithm}[!h]
		\caption{Time-Constrained Reachability with Maximum Probability for PBCNs Using Enhanced $Q$L with TL}
		\label{alg:AOS}
		\renewcommand{\algorithmicrequire}{\textbf{Input:}}
		\renewcommand{\algorithmicensure}{\textbf{Output:}}
		
		\begin{algorithmic}[1]
			\REQUIRE $\omega, \gamma, N, T, \mathcal{X}_{0}, \mathcal{X}_d, Q^*(\mathcal{X}_t,\mathcal{U}_t), \mathcal{X}_{rec}^{'}$  
			\ENSURE $\ Q^{*'}(\mathcal{X}_t,\mathcal{U}_t), \mathcal{X}_{rec}$   
			
			\STATE  Initialize: $Q(\mathcal{X}_{t},\mathcal{U}_{t})\leftarrow Q^*(\mathcal{X}_t,\mathcal{U}_t)$, $\forall \mathcal{X}_{t}\in\mathcal{B}^n,\,\forall \mathcal{U}_{t}\in\mathcal{B}^m$, $\mathcal{X}_{rec}\leftarrow \mathcal{X}_{rec}^{'}$
			
			\FOR{$ep$ = 0, 1, \ldots, $N-1$}
			\STATE $t\leftarrow0,\,\mathcal{X}_t\leftarrow\mathcal{X}_{0}$
			
			\STATE Initialize $\mathcal{X}_{t} = [\mathcal{X}_{0}, 0] $
			\IF {$\mathcal{X}_{t} \,\text{not in the}\, \mathcal{X}_{rec}$}
			\STATE Add $\mathcal{X}_{t}$ into $\mathcal{X}_{rec}$ as a new row
			\STATE Expand $Q$-table by adding one additional row
			\STATE $Q^{*}(\mathcal{X}_{t},\mathcal{U}_{t})\leftarrow Q(\mathcal{X}_{t},\mathcal{U}_{t})$
			\ELSE
			\STATE $Q(\mathcal{X}_{t},\mathcal{U}_{t})\leftarrow Q^{*}(\mathcal{X}_{t},\mathcal{U}_{t})$
			\ENDIF
			
			\WHILE{$(t \leq T)\wedge(\mathcal{X}_{t}\neq\mathcal{X}_{d})\lor(t == 0)$}
			\STATE $\mathcal{U}_t\leftarrow\epsilon\text{-greedy}(\epsilon,\mathcal{X}_t)$
			\STATE $t+1,\mathcal{X}_{t+1},r_{t+1}\leftarrow\operatorname{apply}(\mathcal{U}_t)$
			\STATE $\mathcal{X}_{t+1} = [\mathcal{X}_{t+1}, t+1]$
			
			\IF {$\mathcal{X}_{t+1}\,\text{not in the}\, \mathcal{X}_{rec}$}
			\STATE Add $\mathcal{X}_{t+1}$ into $\mathcal{X}_{rec}$ as a new row
			\STATE Expand $Q$-table by adding one additional row
			\STATE $Q^{*}(\mathcal{X}_{t+1},\mathcal{U}_{t+1})\leftarrow Q(\mathcal{X}_{t+1},\mathcal{U}_{t+1})$
			\ELSE
			\STATE $Q(\mathcal{X}_{t+1},\mathcal{U}_{t+1})\leftarrow Q^{*}(\mathcal{X}_{t+1},\mathcal{U}_{t+1})$
			\ENDIF
			
			\IF {$((t-1) \neq T)\lor(\mathcal{X}_{t} = \mathcal{X}_{d})$}
			\STATE $\begin{aligned}&Q(\mathcal{X}_{t},\mathcal{U}_{t})\leftarrow\alpha_{ep}(r_{t+1}+\gamma\max_{\mathcal{U}}Q(\mathcal{X}_{t+1},\mathcal{U}_{t+1}))+\\&(1-\alpha_{ep})Q(\mathcal{X}_t,\mathcal{U}_{t})\end{aligned}$
			\ELSE
			\STATE
			$\begin{aligned}&Q(\mathcal{X}_{t},\mathcal{U}_{t})\leftarrow\alpha_{ep}\times r_{t+1}+(1-\alpha_{ep})Q(\mathcal{X}_t,\mathcal{U}_{t})\end{aligned}$
			\ENDIF
			
			\ENDWHILE
			\STATE $Q^{*'}(\mathcal{X}_{t},\mathcal{U}_{t})\leftarrow Q(\mathcal{X}_{t},\mathcal{U}_{t})$
			\ENDFOR
			\RETURN $\ Q^{*'}(\mathcal{X}_t,\mathcal{U}_t), \mathcal{X}_{rec}$
		\end{algorithmic}
	\end{algorithm}
	
	\section{Simulation Results}
	\label{sec:guidelines}
	In this section, we give two biologically significant examples to demonstrate the performance of the proposed algorithms. These are a $9$ nodes small-scale PBCN and a $28$ nodes large-scale PBCN.
	\subsection{A Small-scale PBCN}
	\begin{Example}
		Consider the following PBCN\label{119} with $9$ nodes on the lactose operon in the Escherichia Coli, extended from the BCN in\cite{robeva2013mathematical}:
	\end{Example}
	\begin{equation}
		\begin{aligned}
			\mathcal{X}^1(t+1)&=\neg\mathcal{X}^7(t)\wedge\mathcal{X}^3(t); \\
			\mathcal{X}^2(t+1)&=\mathcal{X}^1(t); \\
			\mathcal{X}^3(t+1)&=\neg\mathcal{U}^1(t); \\
			\mathcal{X}^4(t+1)&=\mathcal{X}^5(t)\wedge\mathcal{X}^6(t); \\
			\mathcal{X}^5(t+1)&=\begin{cases}\neg\mathcal{U}^1(t)\wedge\mathcal{X}^2(t)\wedge\mathcal{U}^2(t),\mathsf{P}=0.7;\\\mathcal{X}^5(t),\mathsf{P}=0.3;\end{cases}\\
			\mathcal{X}^6(t+1) &=\mathcal{X}^1(t);\\
			\mathcal{X}^7(t+1)&=\neg\mathcal{X}^4(t)\wedge\neg\mathcal{X}^8(t);\\\mathcal{X}^8(t+1)&=\mathcal{X}^4(t)\vee\mathcal{X}^5(t)\vee\mathcal{X}^9(t); \\
			\mathcal{X}^9(t+1)&=\begin{cases}\neg\mathcal{U}^1(t)\wedge(\mathcal{X}^5(t)\vee\mathcal{U}^2(t)),\mathsf{P}=0.6;\\\mathcal{X}^9(t),\mathsf{P}=0.4,\end{cases}\\
		\end{aligned} \label{222}
	\end{equation}
	where $\mathcal{X}^{i}$ represents the state of the $i$-th node before updating, $\mathcal{U}^{i}$ represents the $i$-th component of the control action and $\mathsf{P}$ refers to the probability. Each node and control action represents the biological significance as shown in Table \ref{11}.
	\begin{table}[!ht] 
		\centering
		\caption{Description of the Specific Meaning of Nodes and Actions}
		\begin{tabular}{|c|c|} \hline 
			Nodes & Explanation  \\ \hline
			$\mathcal{X}^1$ & $M = lac$ mRNA    \\ \hline
			$\mathcal{X}^2$ & $P = lac$ permease \\ \hline
			$\mathcal{X}^3$ & $C = $ catabolite activator protein CAP  \\ \hline
			$\mathcal{X}^4$ & $A = $ high concentration of allolacose (inducer)    \\ \hline
			$\mathcal{X}^5$ & $L = $ high concentration of intracellular lactose    \\ \hline
			$\mathcal{X}^6$ & $B = \beta-\text{galactosidase}$   \\ \hline
			$\mathcal{X}^7$ & $R= $ repressor protein $lac$I    \\ \hline
			$\mathcal{X}^8$ & $A_l = $ (at least) low concentration of intracellular lactose     \\ \hline
			$\mathcal{X}^9$ & $L_l = $ (at least) low concentration of allolacose   \\ \hline
			$\mathcal{U}^1$ & $G_e = \text{input}$   \\ \hline
			$\mathcal{U}^2$ & $L_e = \text{input}$   \\ \hline
		\end{tabular}\label{11}
	\end{table}
	The lac operon model\cite{robeva2013mathematical}, exhibits the correct qualitative behavior by predicting that the operon is activated only when external lactose is present and external glucose is absent. In this case all variables of the model, except for the repressor protein ($\mathcal{X}^7$), are activated. When glucose is available, the operon is inactivated. So we set the desired state ${\mathcal{X}_d}=(1,1,1,1,1,1,0,1,1)$ to study the finite time maximum probability reachability problem.
	
	To evaluate the performance of the algorithms in both the non-TL and TL scenarios, we define the average error over $N$ episodes as follows:
	\begin{equation}
		A_{er}(ep)=\frac{1}{rd_{sum}}\sum_{rd=1}^{rd_{sum}}L_{ep}^{rd},
	\end{equation}
	\begin{equation}
		L_{ep}^{rd}=||Q_{ep}^{rd}(\mathcal{X}_{t},\mathcal{U}_{t})-Q^{*}(\mathcal{X}_{t},\mathcal{U}_{t})||,
	\end{equation}
	where $L_{ep}^{rd}$ represents the loss function computed using $Q$L in the $rd$-th round and $ep$-th episode, $Q_{ep}^{rd}(\mathcal{X}_{t},\mathcal{U}_{t})$ refers to the $Q$-table obtained during the $rd$-th round in the $ep$-th episode. $r$ refers to the total number of repeated experiments.
	\subsubsection{Experiment 1}
	{We unfold the experiment for system \eqref{222} based on Algorithm 1. We study it in two separate scenarios:
		\begin{itemize}
			\item Case 1: $T$ is a constant. The parameters are set as follows: $\omega = $ 0.54, $\beta = $ 0.001, $rd_{sum} = $ 1, $N = 3\times10^6$ episodes and $T = $ 8. The initial state of system \eqref{222} is ${\mathcal{X}_{0}}=(0,0,0,0,0,0,0,0,0)$. The average error results are shown in Figure \ref{fig5}.
			\begin{figure}[!h]
				{\includegraphics[width=\columnwidth]{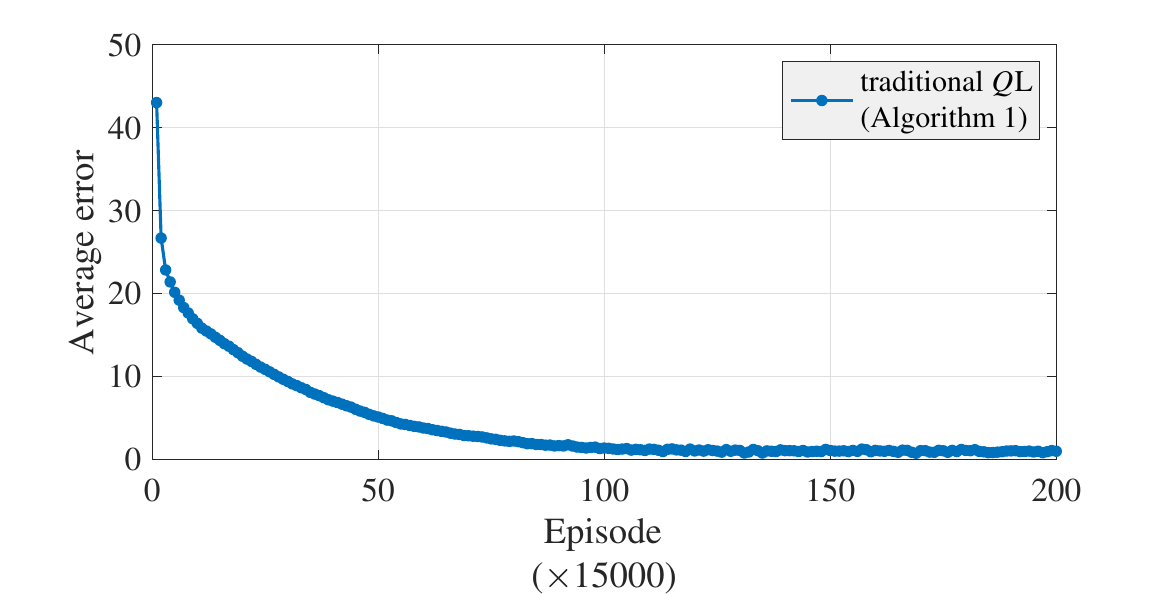}}
				\caption{Performance of Example 1 in Experiment 1 of Case 1 by using Algorithm 1.}
				\label{fig5}
			\end{figure}
			\item Case 2: $T$ follows a  specific distribution. The parameters are set as follows: $\omega = $ 0.54, $\beta = $ 0.0008, $rd_{sum} = $ 1, $N = 4\times10^6$ episodes and $T\sim \mathcal{N}(8,1)$. The initial state of system \eqref{222} is ${\mathcal{X}_{0}}=(0,0,0,0,0,0,0,0,0)$. The average error results are shown in Figure \ref{fig6}.
			\begin{figure}[!h]
				{\includegraphics[width=\columnwidth]{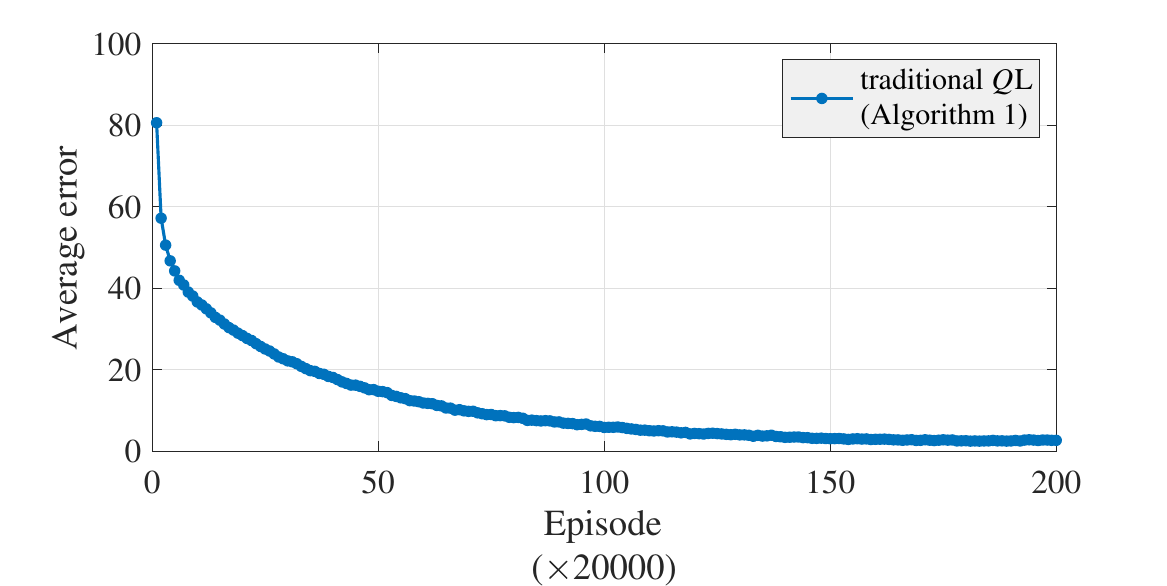}}
				\caption{Performance of Example 1 in Experiment 1 of Case 2 by using Algorithm 1.}
				\label{fig6}
			\end{figure}
		\end{itemize}
		\begin{Remark}
			Here, we select $n'$ = 6 and $n''$ = 2, dividing the domain into $(-\infty, 7), [7, 8), [8, 9), [9, +\infty)$, and then utilize Equations \eqref{1119} for processing.
		\end{Remark}
		\subsubsection{Experiment 2}
		{We unfold the experiment for system \eqref{222} based on Algorithm 2. The parameters are set as follows: $\omega = $ 0.54, $\beta = $ 0.001, $N = 3\times10^6$ episodes, $T = $ 7 and $a = $ 1. The initial state of system \eqref{222} is ${\mathcal{X}_{0}}=(0,0,0,0,0,0,0,0,0)$.} In this experiment, we use the existing information obtained from $T = $ 7 to study the problem $T+a = $ 8 in three scenarios, namely, without using TL, using TL method 1 (TL1), and using TL method 2 (TL2), respectively. For each scenario, we conduct $rd_{sum} =$ 25 repetitions of the experiment separately, and plot the average error in Figure \ref{fig4}. Analyzing the results reveals that all three algorithms converge and combining the TL methods leads to a faster convergence of the algorithm. Moreover, TL2 outperforms TL1.
		\begin{Remark}
			We count the total number of appearances for each visited state during the $30\times10^5$ episodes and find that the states corresponding to the following rows ($1032$, $1095$, $1539$, $1540$, $1601$, $1602$, $1603$, $1607$, $2052$, $2056$, $2114$, $2308$, $2370$) of the $Q$-table are visited less than $1\times10^5$ times. Since the number of visits is less than 0.03 times the total number of episodes, we don't consider it meaningful for our study and therefore omit the calculation of the average error for the aforementioned states.
		\end{Remark}
		\begin{figure}[!h]
			{\includegraphics[width=\columnwidth]{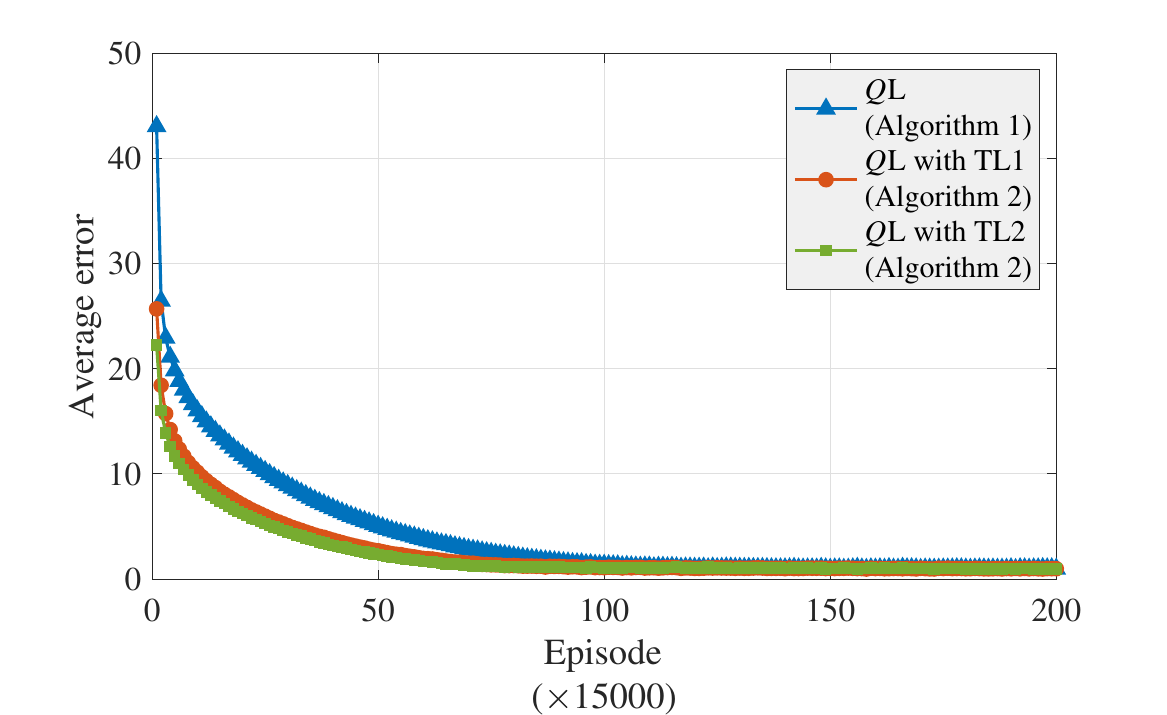}}
			\caption{Performance of Example 1 in Experiment 2 by using Algorithm 2.}
			\label{fig4}
		\end{figure}
		\subsection{A Large-scale PBCN}
		\begin{Example}
			Consider the following PBCN\label{1116} with $28$ nodes, which is a reduced-order model of 32-gene T-cell receptor kinetics model, extended from the BCN in\cite{zhang2020efficient}:	
			\begin{gather}
				\mathcal{X}^1(t+1)=\mathcal{X}^6(t)\wedge\mathcal{X}^{13}(t);\, \mathcal{X}^2(t+1) =\mathcal{X}^{25}(t);\, \mathcal{X}^3(t+1) \notag
				\\
				=\mathcal{X}^2(t);\, \mathcal{X}^4(t+1) =\mathcal{X}^{28}(t);\, \mathcal{X}^5(t+1) =\mathcal{X}^{21}(t);\, \mathcal{X}^6(t+ \notag
				\\
				1) =\mathcal{X}^5(t);\, \mathcal{X}^{7}(t+1) =(\mathcal{X}^{15}(t)\wedge\mathcal{U}^{2}(t))\vee(\mathcal{X}^{26}(t)\wedge\mathcal{U}^{2}(t)); \notag
				\\
				\mathcal{X}^{8}(t+1)=\mathcal{X}^{14}(t);\, \mathcal{X}^{9}(t+1)=\mathcal{X}^{18}(t);\, \mathcal{X}^{10}(t+1)=\mathcal{X}^{25}(t) \notag
				\\
				\wedge\mathcal{X}^{28}(t);\,
				\mathcal{X}^{11}(t+1)=\neg\mathcal{X}^{9}(t);\,
				\mathcal{X}^{12}(t+1)=\mathcal{X}^{24}(t);\, \mathcal{X}^{13}(t+ \notag
				\\
				1)=\mathcal{X}^{12}(t);\,
				\mathcal{X}^{14}(t+1)=\mathcal{X}^{28}(t);\,
				\mathcal{X}^{15}(t+1)=(\neg\mathcal{X}^{20}(t))\wedge \notag
				\\
				\mathcal{U}^{1}(t)\wedge\mathcal{U}^{2}(t);\,
				\mathcal{X}^{16}(t+1)=\mathcal{X}^{3}(t); \,
				\mathcal{X}^{17}(t+1)=\neg\mathcal{X}^{11}(t); \notag
				\\
				\mathcal{X}^{18}(t+1)=\mathcal{X}^{2}(t);\,
				\mathcal{X}^{19}(t+1)=(\mathcal{X}^{10}(t)\mathcal{X}^{11}(t)\wedge\mathcal{X}^{25}(t)\wedge \notag
				\\
				\mathcal{X}^{28}(t))\vee(\mathcal{X}^{11}(t)\wedge\mathcal{X}^{23}(t)\wedge\mathcal{X}^{25}(t)\wedge\mathcal{X}^{28}(t));\,
				\mathcal{X}^{20}(t+1)= \notag
				\\
				\mathcal{X}^{7}(t)\vee\neg\mathcal{X}^{26}(t); \,
				\mathcal{X}^{21}(t+1)=	\mathcal{X}^{11}(t)\vee\mathcal{X}^{22}(t);\,
				\mathcal{X}^{22}(t+1) \notag
				\\
				=\mathcal{X}^{2}(t)\wedge\mathcal{X}^{18}(t);\,
				\mathcal{X}^{23}(t+1)=\mathcal{X}^{15}(t); \,
				\mathcal{X}^{24}(t+1)=\mathcal{X}^{18}(t); \notag
				\\
				\mathcal{X}^{25}(t+1)=\mathcal{X}^{8}(t);\,
				\mathcal{X}^{26}(t+1)=\neg\mathcal{X}^{4}(t)\wedge\mathcal{U}^{3}(t),~\mathsf{P}~=0.5;, \notag
				\\
				\mathsf{and}~\mathcal{X}^{26}(t+1)=\mathcal{X}^{26}(t),~\mathsf{P}~=~0.5;
				\mathcal{X}^{27}(t+1)=\mathcal{X}^{7}(t)\lor \notag
				\\
				(\mathcal{X}^{15}(t)\land\mathcal{X}^{26}(t)); \,
				\mathcal{X}^{28}(t+1)=\neg\mathcal{X}^4(t)\wedge\mathcal{X}^{15}(t)\wedge\mathcal{X}^{27}(t),\label{1121}
			\end{gather}
			where each node and action represents the biological significance as shown in Table \ref{1111}. 
			
			Studying the reachability problem in T-cell receptor kinetics models plays a crucial role in immunotherapy research, such as optimizing the design of immunotherapy strategies and enabling individualized treatment decisions. We set the desired state ${\mathcal{X}_d}=(0,0,0,0,1,1,1,0,0,0,1,0,0,0,0,0,0,0,0,1,1,\\0,0,0,0,1,1,0)$ to study the finite time maximum probability reachability problem.
		\end{Example}
		\begin{table}[!ht] 
			\centering
			\caption{Description of the Specific Meaning of Nodes and Actions}
			\begin{tabular}{|c|c||c|c|} \hline 
				Nodes & Explanation & Nodes & Explanation \\ \hline
				$\mathcal{X}^1$ & AP1 & $\mathcal{X}^{16}$ & NFAT \\ \hline
				$\mathcal{X}^2$ & Ca/ DAG & $\mathcal{X}^{17}$ & NFkB  \\ \hline
				$\mathcal{X}^3$ & Calcin & $\mathcal{X}^{18}$ & PKCth \\ \hline
				$\mathcal{X}^4$ & cCbl & $\mathcal{X}^{19}$ & PKCg(act) \\ \hline
				$\mathcal{X}^5$ & ERK & $\mathcal{X}^{20}$ & PAGCsk \\ \hline
				$\mathcal{X}^6$ & Fos & $\mathcal{X}^{21}$ & MEK/ Ras \\ \hline
				$\mathcal{X}^7$ & Fyn & $\mathcal{X}^{22}$ & RasGRP1 \\ \hline
				$\mathcal{X}^8$ & Gads & $\mathcal{X}^{23}$ & Rlk \\ \hline
				$\mathcal{X}^9$ & IKKbeta & $\mathcal{X}^{24}$ & SEK \\ \hline
				$\mathcal{X}^{10}$ & Itk & $\mathcal{X}^{25}$ & IP3/ SLP76 \\ \hline
				$\mathcal{X}^{11}$ & Grb2Sos/ IkB/ PLCg(bind) & $\mathcal{X}^{26}$ & TCRbind \\ \hline
				$\mathcal{X}^{12}$ & JNK & $\mathcal{X}^{27}$ & TCRphos \\ \hline
				$\mathcal{X}^{13}$ & Jun & $\mathcal{X}^{28}$ & ZAP70 \\ \hline
				$\mathcal{X}^{14}$ & LAT & $\mathcal{U}^{1}$ & CD8 \\ \hline
				$\mathcal{X}^{15}$ & Lck & $\mathcal{U}^{2}$ & CD45 \\ \hline
				& & $\mathcal{U}^{3}$ & TCRlig \\ \hline
			\end{tabular} \label{1111}
		\end{table}
		\subsubsection{Experiment 3}
		{We unfold the experiment for system \eqref{1121} based on Algorithm 3. We study it in two separate scenarios:
			\begin{itemize}
				\item Case 1: $T$ is a constant. The parameters are set as follows: $\omega = $ 0.54, $N = 2\times10^6$ episodes and $T = $ 9. The initial state of for system \eqref{1121} is ${\mathcal{X}_{0}} = (0,0,0,0,1,1,0,0,0,0,1,0,1,0,0,0,1,0,0,1,0,0,0,0,\\0,0,0,0)$.
				The average rewards in the traing process by Algorithm 3 are shown in Figure \ref{fig7}.
				
				\item Case 2: $T$ follows a  specific distribution. The parameters are set as follows: $\omega = $ 0.54, $N = 2.5\times10^6$ episodes and $T\sim \mathcal{N}(10,1)$. The initial state of for system \eqref{1121} is ${\mathcal{X}_{0}}=(0,0,0,0,1,1,0,0,0,0,1,0,1,0,0,0,1,0,0,1,0,0,0,0,\\0,0,0,0)$. The average rewards in the training process by Algorithm 3 are shown in Figure \ref{fig8}.
				
				\begin{figure}[!h]
					{\includegraphics[width=\columnwidth]{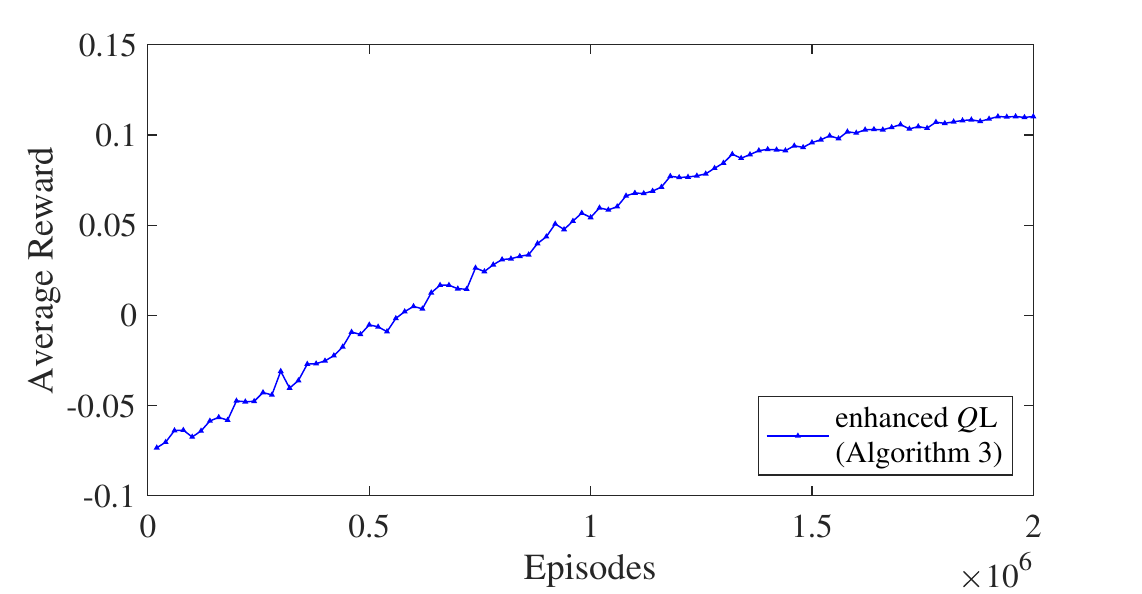}}
					\caption{Average rewards during training by using Algorithm 3 of Example 2 in Experiment 3 of Case 1.}
					\label{fig7}
				\end{figure}
				\begin{figure}[!h]
					{\includegraphics[width=\columnwidth]{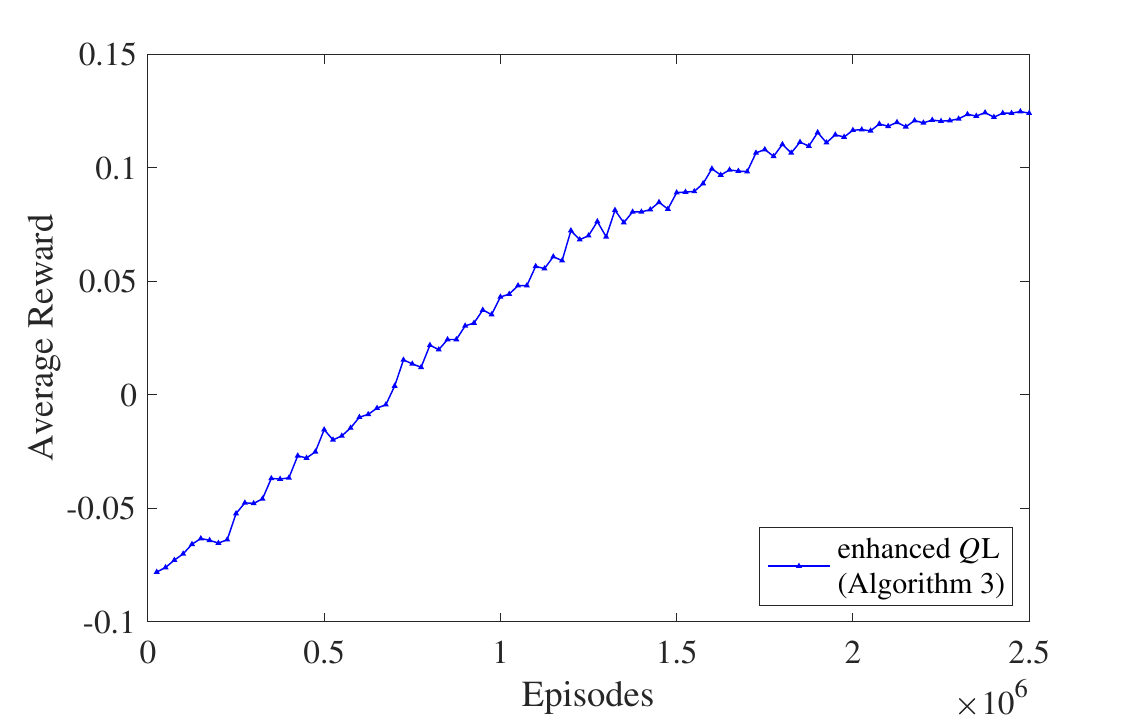}}
					\caption{Average rewards during training by using Algorithm 3 of Example 2 in Experiment 3 of Case 2.}
					\label{fig8}
				\end{figure}
			\end{itemize}
			
			The term ``average rewards" refers to taking the average of rewards over the neighboring 1000 episodes. This approach helps to reduce the impact of initial states on rewards and provides a more objective representation of the training's impact on rewards.
			\begin{Remark}
				Here, we select $n'$ = 8 and $n''$ = 2, dividing the domain into $(-\infty, 9), [9, 10), [10, 11), [11, +\infty)$, and then utilize Equations \eqref{1119} for processing.
			\end{Remark}
			\subsubsection{Experiment 4}
			{We unfold the experiment  for system \eqref{1121} based on Algorithm 4. The parameters are set as follows: $\omega = $ 0.54, $rd_{sum} = $1, $N = 2\times10^6$ episodes, $T = $ 9 and $a = $ 1. The initial state of system \eqref{1121} is ${\mathcal{X}_{0}}=(0,0,0,0,1,1,0,0,0,0,1,0,1,0,0,0,1,0,0,1,0,0,0,0,0,0,\\0,0)$. In this experiment, we use the knowledge obtained from $T = $ 9 to study the problem $T+a = $ 10.} We compare  algorithm's performance between scenarios without using TL and with TL. The average error is used as the metric for comparison. As shown in Figure \ref{fig10}, the results indicate that combining TL methods leads to better convergence of the algorithm. Figure \ref{fig9} demonstrates that the agent can reach the T-cell receptor kinetics model at $\mathcal{X}_d$ with a probability of 1 in a minimum of 9 steps.
			\begin{Remark}
				We consider the states that have been visited fewer than or equal to $0.05 \times N$ times to be insignificant for our study. Therefore, we do not calculate the average error for these corresponding states.
			\end{Remark}
			\begin{figure}[!h]
				{\includegraphics[width=\columnwidth]{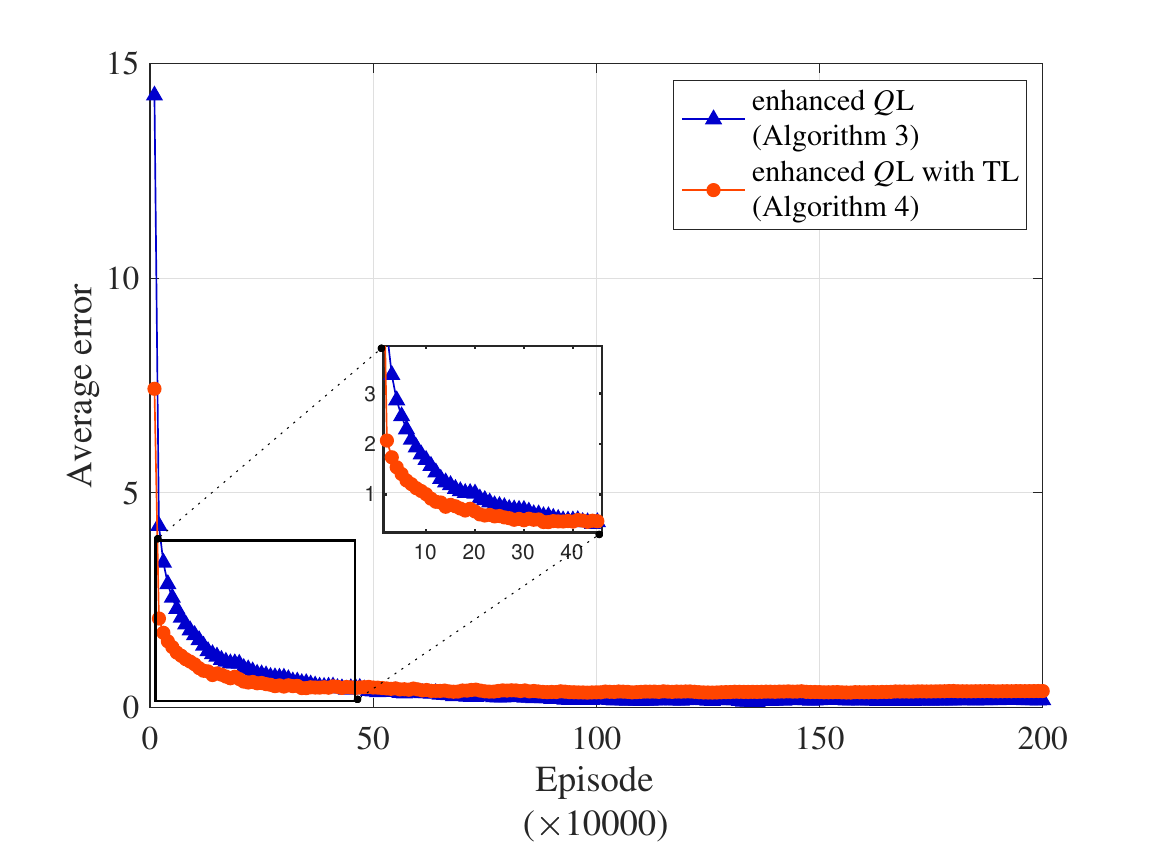}}
				\caption{Performance of Example 2 in Experiment 2 by using Algorithm 4.}
				\label{fig10}
			\end{figure}
			\begin{figure}[!h]
				{\includegraphics[width=\columnwidth]{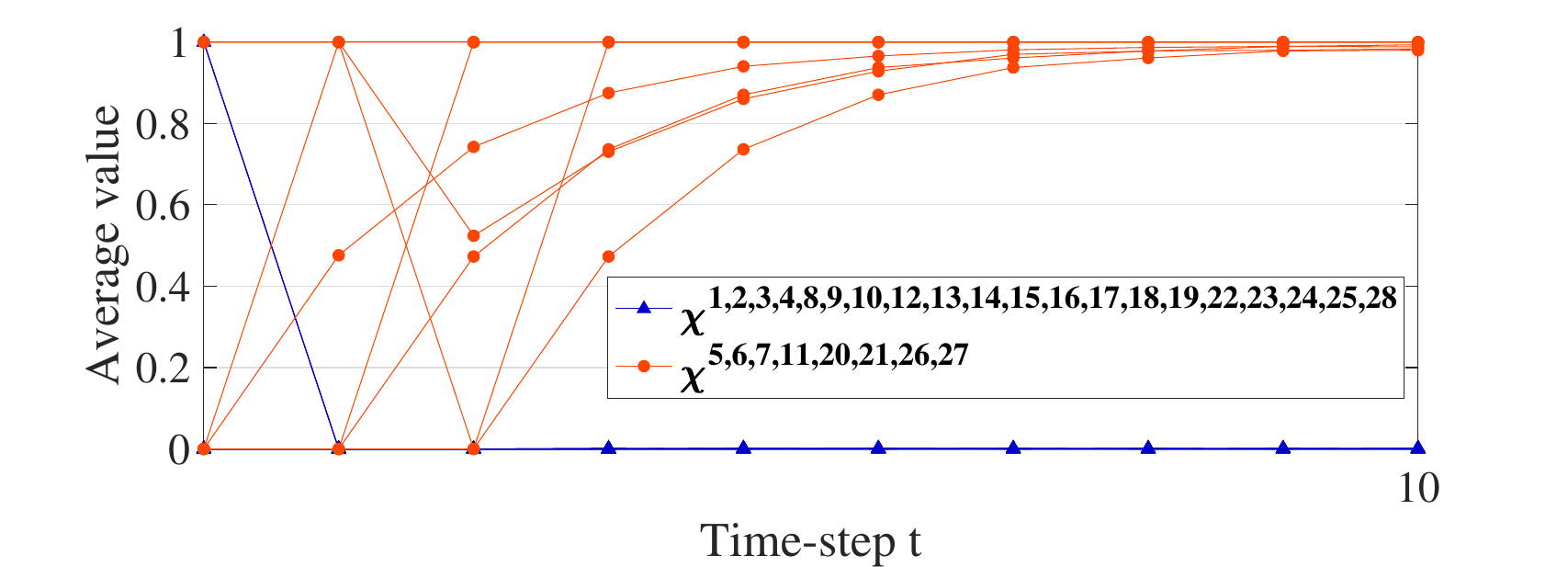}}
				\caption{The optimal policy of the T-cell receptor kinetics to achieve the reachability by using Algorithm 4 of Example 2 in Experiment 4.}
				\label{fig9}
			\end{figure}
			\subsection{Details}
			\begin{itemize}
				\item In the Algorithms 1 and 2, we set the learning rate as a generalized harmonic series $\alpha_{ep}=\min\{1,\frac{1}{(\beta* ep)^\omega}\}$. In the Algorithms 3 and 4, we set the learning rate as a generalized harmonic series $\alpha_{ep}=\min\{1,\frac{3}{(ep+1)^\omega}\}$. Both of them satisfy condition 2) in Lemma 1.
				\item In all the proposed algorithms, the greedy rate $\epsilon$ is defined as $-0.99/N*ep+1$, which changes with a linear decaying value, from 1 to 0.01.
			\end{itemize}
			\section{Conclusion}
			In this paper, we investigate the problem of maximizing the probability of reachability in PBCNs within the finite time frame using a reinforcement learning-based approach. Our research considers a wider range of finite-time scenarios, including cases where $T$ is subject to a dynamic framework, such as following a certain distribution. Moreover, we introduce TL by leveraging prior knowledge to accelerate convergence speed and propose an enhanced $Q$L technique to effectively handle the reachability problems in large-scale PBCNs, resulting in the development of Algorithm 3 and Algorithm 4. Finally, we provide two biologically meaningful network examples to demonstrate the effectiveness of our approach.

			\bibliographystyle{IEEEtran}
			\bibliography{kauffman.bib}
			
		\end{document}